\documentclass[apj]{emulateapj}

\def\Journal#1#2#3#4{{#4}, {#1}, {#2}, #3}

\def\AAA{A\&A}
\def\ApJ{ApJ}

\def\Aph{Astropart. Phys.}
\def\ApJS{ApJSS}
\def\ML{Machine Learning}

\def\NIMA{Nucl. Instrum. Methods A}

\begin{document}

\title{Discovery of Very High Energy $\gamma$-Rays from Markarian~180 Triggered by an Optical Outburst}

%
\author{
 J.~Albert\altaffilmark{a}, 
 E.~Aliu\altaffilmark{b}, 
 H.~Anderhub\altaffilmark{c}, 
 P.~Antoranz\altaffilmark{d}, 
 A.~Armada\altaffilmark{b}, 
 M.~Asensio\altaffilmark{d}, 
 C.~Baixeras\altaffilmark{e}, 
 J.~A.~Barrio\altaffilmark{d}, 
 H.~Bartko\altaffilmark{g}, 
 D.~Bastieri\altaffilmark{h}
% ...
}
 \altaffiltext{a} {Universit\"at W\"urzburg, D-97074 W\"urzburg, Germany}
 \altaffiltext{b} {Institut de F\'\i sica d'Altes Energies,  
E-08193 Bellaterra, Spain}
 \altaffiltext{c} {ETH Zurich, CH-8093 Switzerland}
 \altaffiltext{d} {Universidad Complutense, E-28040 Madrid, Spain}
 \altaffiltext{e} {Universitat Aut\`onoma de Barcelona, E-08193 Bellaterra, Spain}
% ...
\altaffiltext{g} {Max-Planck-Institut f\"ur Physik, D-80805 M\"unchen, Germany}
 \altaffiltext{h} {Universit\`a di Padova and INFN, I-35131 Padova, Italy} 
% ...


\begin{abstract}

The high-frequency--peaked BL Lacertae object Markarian~180 (Mrk~180) was
observed to have an optical outburst in 2006 March, triggering a Target of
Opportunity observation with the MAGIC telescope. The source was observed for
12.4 hr, and very high energy $\gamma$-ray emission was detected with a
significance of 5.5~$\sigma$.  An integral flux above 200~GeV of
$(2.3\pm0.7)\times10^{-11}\, \mbox{cm}^{-2}\,\mbox{s}^{-1}$ was measured,
corresponding to 11\% of the Crab Nebula flux. A rather soft spectrum with a
photon index of $-3.3\pm0.7$ has been determined. No significant flux
variation was found.

\end{abstract}



\section{Introduction}

The search for very high energy (VHE, defined as $E\gtrsim 100$~GeV)
$\gamma$-ray emission from active galactic nuclei (AGNs) is one of the major
goals for ground-based $\gamma$-ray astronomy. New detections open the
possibility of a phenomenological study of the physics inside the relativistic
jets in AGNs, in particular, to understand both the origin of the VHE
$\gamma$-rays as well as the correlations between photons of different energy
ranges (from radio to VHE). The number of reported VHE $\gamma$-ray--emitting
AGNs has been slowly increasing and is currently 12 (2006 June).  Six of
them have been seen by MAGIC: Mrk~421 \citep{magic421}, Mrk~501
\citep{mazin}, 1ES1959+650 \citep{magic1959}, 1ES2344+514 \citep{mazin},
1ES1218+304 \citep{magic1218}, and PG1553+113 \citep{magic1553}.

The known VHE $\gamma$-ray--emitting AGNs are variable in flux in all wave bands.
Correlations between X-ray and $\gamma$-ray emission have been found (e.g.,
\citet{fossati04}), although the relationship has proven to be rather
complicated, with $\gamma$-ray flares also being detected in the absence of
X-ray flares \citep{holder,krawczynski} and vice versa \citep{rebillot}. The
optical-TeV correlation has yet to be studied, but the optical-GeV correlations
seen in 3C~279 \citep{hartman01} suggest that at least in some sources, such
correlations exist. Using this as a guideline, the MAGIC collaboration has been
performing Target of Opportunity observations whenever alerted that
sources were in a high flux state in the optical and/or X-ray band.

The AGN Mrk~180 (1ES 1133+704) is a well-known high-frequency--peaked BL Lac
(HBL) object at a redshift of $z=0.045$ \citep{falco}. The spectral energy
distribution (SED) of HBL objects exhibits a generic two-bump structure: one peak with
a maximum in the X-ray band and the other peak located in the GeV-TeV band.
The radiation is produced in a highly beamed plasma jet, which is almost
aligned with the observer's line of sight. A double-peaked SED is normally
attributed to a population of relativistic electrons, where one peak is due to
synchrotron emission in the magnetic field of the jet and where the second peak is
caused by inverse Compton (IC) scattering of low-energy photons. The low-energy
photons can be external to the jet (external Compton scattering; \citet{dermer}) or
are produced within the jet via synchrotron radiation (synchrotron
self-Compton [SSC] scattering; \citet{maraschi}). Models based on the acceleration of
hadrons can also sufficiently describe the observed SEDs and light curves
\citep{mannheim91,massaro2}. All of the up to now known AGNs with strong GeV/TeV
$\gamma$-emission belong to the HBL object class.

In many cases, the second peak of the SED is not observable because of low 
sensitivity above a few times 100~MeV of satellite-borne detectors and a
too high energy threshold of ground-based $\gamma$-ray detectors. In case 
of Mrk~180, HEGRA \citep{54AGN} and Whipple \citep{horan}
observed this object but were only able to derive flux upper limits, and EGRET
did not detect the source \citep{fichtel, hartman}.

In the optical, Mrk~180 is characterized by a bright host galaxy
$R=14.17\pm0.02$ mag and a much fainter (variable) core $R=15.79\pm0.02$ mag
\citep{nilsson}. The  variations in total optical brightness are therefore
small. The source is observed regularly as part of the Tuorla Observatory
Blazar monitoring program\footnote{See http://users.utu.fi/kani/1m/.} with the Tuorla 1~m
telescope and the KVA 35~cm telescope\footnote{See http://tur3.tur.iac.es/.}.  To determine
the core flux, we had to subtract the flux of the host galaxy and a nearby star
within the $5''$ aperture radius (together R=14.96 mag, or 3.2 mJy; \linebreak
Nilsson, private communication).

After Mrk~180 had been detected in the X-rays by {\it HEAO-1} \citep{HEAO}, it
was observed by various satellites with fluxes ranging from 6.3 to $22\times
10^{-12}$ ergs cm$^{-2}$ s$^{-1}$ \citep{donato} at around 1~keV and from $5.0$ to
$9.8\times10^{-12}$ ergs cm$^{-2}$ s$^{-1}$ in the 2--10~keV band
\citep{perlman,donato05}. The source is monitored by the all-sky monitor (ASM) 
on board the {\it Rossi X-Ray Timing Explorer}.

In this Letter we present the first detection of VHE $\gamma$-ray
emission from Mrk~180.


\section{Observations and Data Analysis}

The MAGIC telescope (Major Atmospheric Gamma Imaging
Cherenkov telescope; \citet{magic}) is located on the Canary Island La
Palma (2200 m above sea level, 28\fdg45$'$ north, 17\fdg54$'$ west).  
MAGIC is currently the largest imaging atmospheric Cherenkov telescope 
with a 17~m--diameter tessellated reflector dish \citep{cortina}. The
3\fdg5 field of view camera comprises 576 photomultipliers (PMTs) with
enhanced quantum efficiency.  The accessible energy range spans from 50--60 GeV
(trigger threshold at small zenith angles) up to tens of TeV.  The $\gamma$
point-spread function is about 0\fdg1.

\begin{figure}[t]
\begin{center}
\begin{minipage}[b]{0.45\textwidth}
\includegraphics*[width=0.9\textwidth,clip]{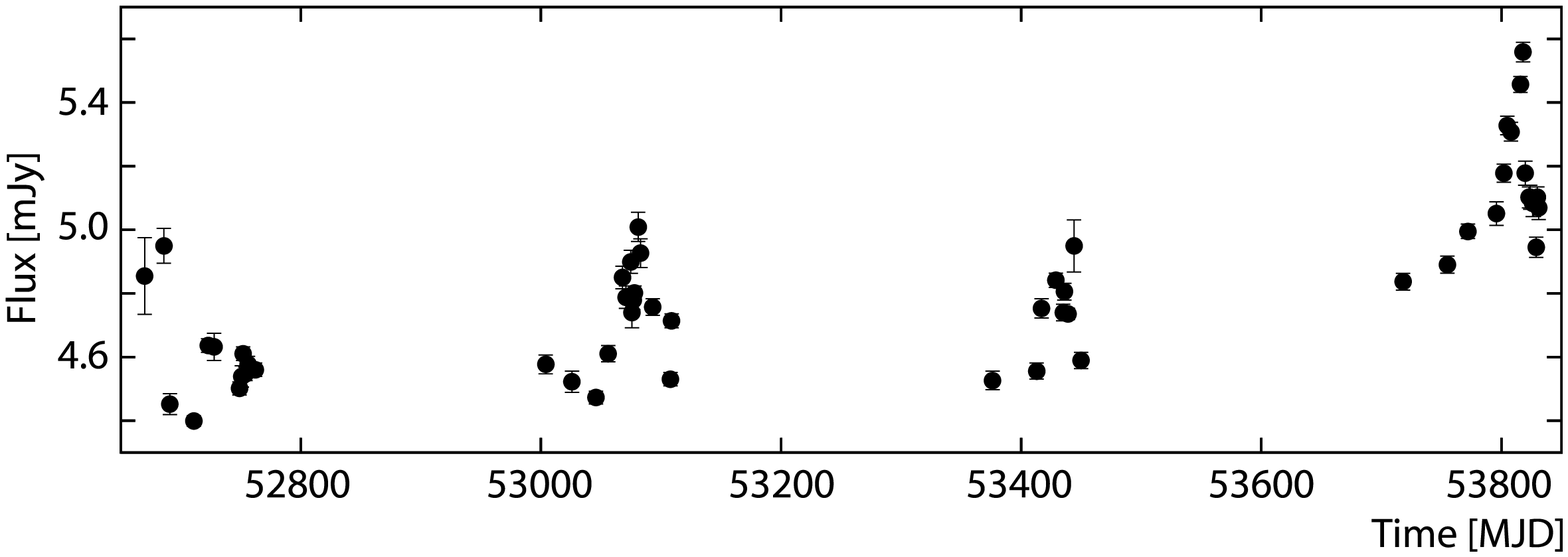}
\caption{\label{fig_opt} {\it R}-band light curve of Mrk~180 extending
  from 2003 January to the end of 2006 March as measured by the KVA telescope; 
  5 mJy is equivalent to 14.47 mag.}
\end{minipage}
\end{center}
\end{figure}

\begin{figure}[t]
\begin{center}
\begin{minipage}[b]{0.45\textwidth}
\includegraphics*[width=0.9\textwidth,clip]{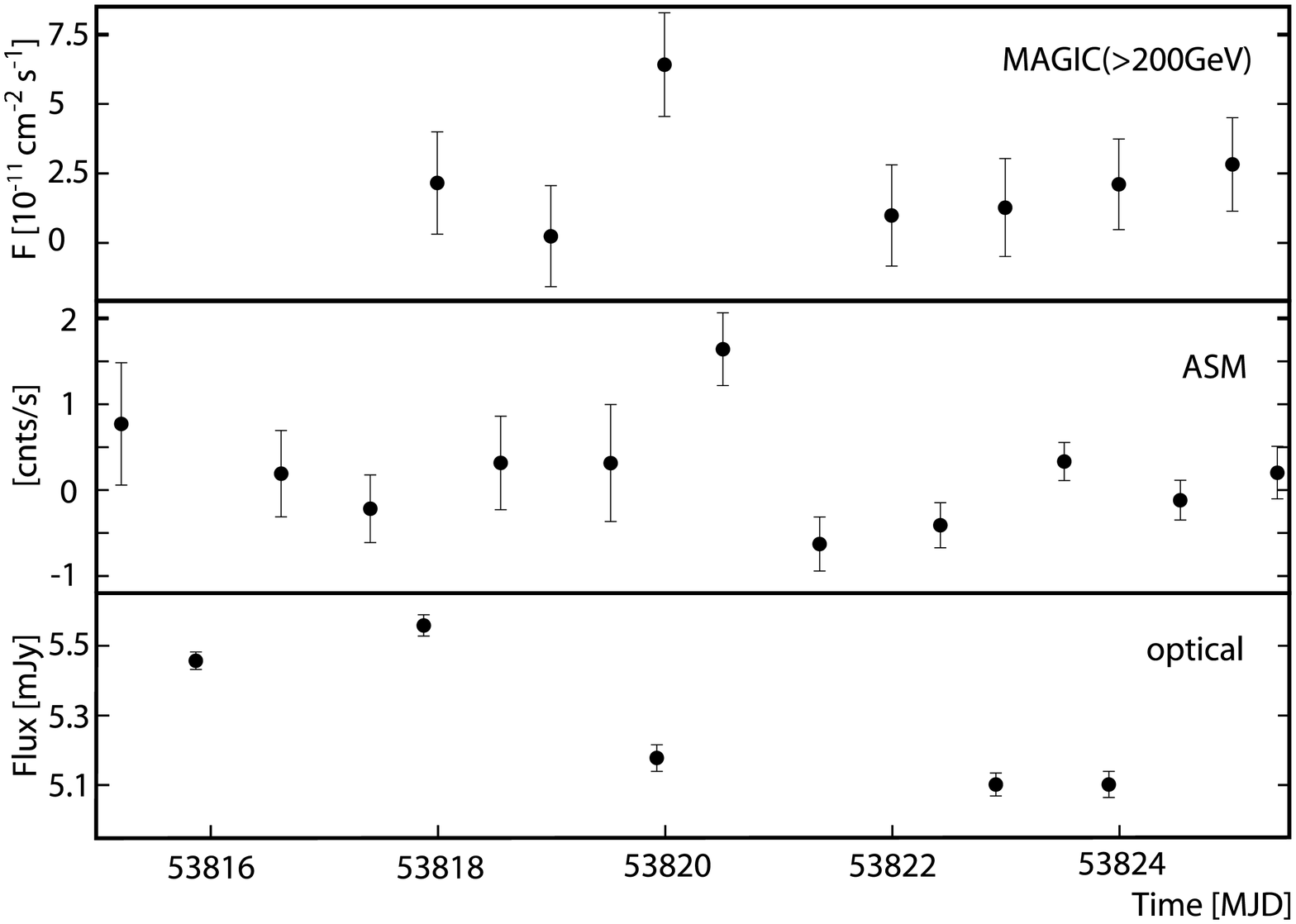}
\caption{\label{fig_LC} Light curve of Mrk~180 for 
    MJD = 53815--53825 (March 21--31).
    {\it Upper panel}: VHE $\gamma$-rays measured by MAGIC above 200~GeV.
    {\it Middle panel}: ASM count rate.
    {\it Lower panel}: Optical flux measured by KVA.} 
\end{minipage}
\end{center}
\end{figure}

The observation of Mrk~180 was triggered by a brightening of the source in the
optical on 2006 March 23.  The alert was given as the core flux increased by
50\% from its quiescent level value as determined from over 3 years of data
recording, shown in Fig.~\ref{fig_opt}.  During the MAGIC observation, optical
follow-up observations were performed with KVA. The simultaneous MAGIC, ASM, and
KVA light curve is shown in Fig.~\ref{fig_LC}.  Around this time, Mrk 180 was
also observed as part of the AGN-monitoring program by the University of
Michigan Radio Observatory (UMRAO). 
No evidence of flaring was found between 2006 January and April.

Mrk~180 was observed in 2006 during 8 nights (from March 23 to 31), for a total
of 12.4 hr,  at zenith angles ranging from $39^{\circ}$ to $44^{\circ}$.
The observations were performed in the so-called wobble mode \citep{daum}, in
which the telescope is pointed alternatively for 20 minutes to two opposite sky
positions at 0\fdg4 off the source.  Runs with unusual trigger rates due
to detector problems or adverse atmospheric conditions were rejected.  The
total observation time was thus reduced to 11.1 hr.

The data were analyzed using the standard analysis and calibration programs for
the MAGIC telescope \citep{bretz,rwagner,gaug}.  For the $\gamma$/hadron shower
separation a multidimensional classification technique based on the Random
Forest method was used \citep{breiman,rf}.  The cuts for the $\gamma$/hadron
separation were trained using a fraction of randomly chosen data to represent
the background (hadrons) and were compared to Monte Carlo (MC) $\gamma$ events with
the same zenith angle distribution (CORSIKA ver. 6.023;
\citet{corsika,majumdar}).  The cuts were then chosen such that the overall cut
efficiency for MC $\gamma$ events was about 50$\%$.  The number of excess
events is determined as the difference between the source and background region
in the $\theta^{2}$ distributions, with $\theta$ being the angular distance between
the source position in the sky and the reconstructed arrival position of the
air shower. The latter position is determined for each shower image by means of
the so-called DISP method \citep{fomin,lessard,domingo}, using image shape
parameters \citep{hillas}.  In order to determine the background distribution,
three background regions of the same size are chosen symmetrically to the
on-source region with respect to the camera center.  A final cut of
$\theta^2<0.024$ is applied to determine the significance of the signal and the
number of excess events.  The energy of the $\gamma$-ray candidates was also
estimated using the Random Forest technique.  The applied cuts were chosen to
be looser than the ones in Fig.~\ref{fig_theta2} in order to gain statistics on
the $\gamma$-ray candidates. Due to the large zenith angle of the observation,
the corresponding energy threshold (defined as the peak in the differential
energy distribution of the MC $\gamma$ events) after cuts was about 200~GeV.
Effects on the spectrum determination, introduced by the limited energy
resolution of the detector, were corrected using the ``unfolding'' methods according to
\citet{anykeyev}.


\section{Results}

\begin{figure}[t]
\begin{center}
\begin{minipage}[b]{0.45\textwidth}
\includegraphics*[width=1\textwidth,angle=0,clip]{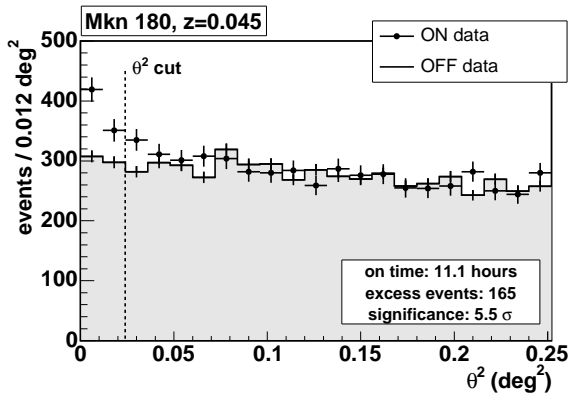}
\caption{\label{fig_theta2} The $\theta^2$-distribution for the   
    ON-source data ({\it filled circles}) and normalized OFF-source events
    ({\it gray histogram}) for Mrk~180.
    The vertical dotted line indicates the 
    $\theta^2$ cut used to determine excess events. The total excess
    of 165 events corresponds to a significance of 5.5~$\sigma$.}
\end{minipage}
\end{center}
\end{figure}

The distribution of $\theta^2$-values after cuts is shown in
Fig.~\ref{fig_theta2}.  The signal of 165 events over 605.2 normalized
background events corresponds to a 5.5~$\sigma$ excess using equation (17) in
\citet{LiMa}. The shape of the excess is consistent with a pointlike source
seen by MAGIC.

No evidence of flux variability was found.  The fit to the nightly integrated
flux is consistent with a constant emission ($\chi^{2}=7.1$, 6 degrees of freedom).
Fig.~\ref{fig_LC} shows the VHE light curve together with the ASM daily averages
and the {\it R}-band flux data. The X-ray flux of the source is generally below the
ASM sensitivity, but on March 25 a 3~$\sigma$ excess was observed, which
suggests that the source was also active in X-rays.  The optical flux reached
its maximum on the night MAGIC started the observations (March 23) and began to
decrease afterward.

\begin{figure}[t]
\begin{center}
\begin{minipage}[b]{0.45\textwidth}
\includegraphics*[width=1\textwidth,angle=0,clip]{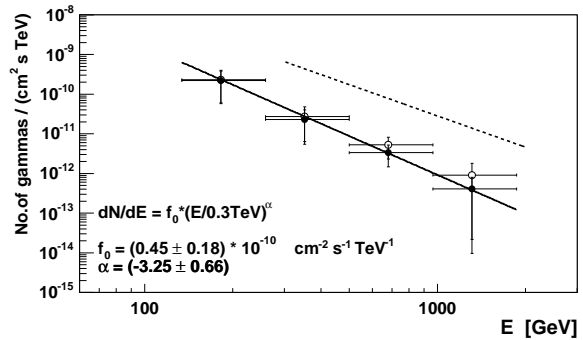}
\caption{\label{fig_spectrum} Differential energy spectrum of Mrk~180.
    {\it Full circles}: Spectrum measured by MAGIC. 
    {\it Open circles}: De-absorbed energy spectrum (see text).
    The horizontal bars indicate the size of each energy bin.
    The black line represents a power-law fit to the
    measured spectrum. The fit parameters are listed in the figure.
    For comparison, the Crab Nebula energy spectrum as derived from MAGIC 
    data \citep{rwagner} is shown ({\it dotted line}).}
\end{minipage}
\end{center}
\end{figure}

The measured energy spectrum of Mrk~180 is shown in Fig.~\ref{fig_spectrum}.
Assuming a power-law spectrum, we obtained the following parameterization: \[
{\frac{\mathrm{d}N}{\mathrm{d}E}}=(4.5\pm1.8)\times10^{-11}\left({\frac{E}{0.3\,\mathrm{TeV}}}\right)^{-3.3\pm0.7}\,
\frac 1 {\mbox{TeV}\,\mbox{cm}^{2}\,\mbox{s}}  \] The observed integral flux
above 200~GeV is $F(E>200\:\mathrm{GeV})=(2.25\pm0.69)\times10^{-11}
\mbox{cm}^{-2}\mbox{s}^{-1}$, which corresponds to $1.27\times10^{-11}\,
\mathrm{ergs} \mathrm{cm}^{-2} \mathrm{s}^{-1} $, or 11\% of the Crab Nebula
flux measured by MAGIC.  The errors are statistical only.  We estimate the
systematic errors to be around 50\% for the absolute flux level and 0.2 for the
spectral index.  The large systematic flux error is a consequence of the steep
slope.  The main error contributions  are the uncertainty in the atmospheric
transmission, the reflectivity and focusing uncertainty of the mirrors and
light catchers in front of the PMTs, and the uncertainty in the effective quantum
efficiency of the PMT and in the photoelectron-to-signal conversion. A second
independent analysis gave results in very good agreement with the quoted
numbers.

The VHE $\gamma$-rays from Mrk~180 are partially absorbed by the low-energy
photons of the evolving extragalactic background light (EBL; see
\citet{nikishov,gould,stecker,HauserDwek}).  Given the redshift z=0.045 of
Mrk~180, the effect is small for photons with energies below 1 TeV. We calculate
the optical depth and the resulting attenuation of the VHE $\gamma$-rays from
Mrk~180 using the number density of the evolving EBL provided by the best-fit
model of \citet{kneiske}.  This state-of-the-art model is consistent with the
recently derived upper limits on the EBL inferred from arguments about AGN spectra
\citep{hessebl}.  The attenuation was determined by the numerical integration of
equation (2) from \citet{dwek}.  The de-absorbed energy spectrum of Mrk~180 is
also shown in Fig.~\ref{fig_spectrum} ({\it open circles}).  A fit with a simple
power-law to the de-absorbed spectrum reveals a slope with $\alpha=-2.8\pm0.7$.


\section{Discussion and conclusions}

In this Letter we have presented the first detection of VHE $\gamma$-ray
emission from Mrk~180. The discovery was triggered by an optical flare, but no
significant variations in the VHE regime were found. The short observation
period and the small signal prevent us from carrying out detailed studies. It is
therefore impossible to judge whether the detected VHE flux level represents
a flaring or a quiescent state of the AGN.

Earlier observations by other experiments have only set upper limits on the VHE
$\gamma$-ray emission from Mrk~180. HEGRA observed the source for 9.8 hr,
resulting in an upper limit (99\% c.l.) at 1.5~TeV of 12\% of the Crab Nebula
flux \citep{54AGN}.  Whipple observed Mrk~180 during three observation periods
for a total of 26.8 hr, resulting in a flux upper limit at 300~GeV of 10.8\%
of the Crab Nebula flux \citep{horan}.  Both upper limits are above the flux
presented in this Letter. 

\begin{figure}[t]
\begin{center}
\begin{minipage}[b]{0.45\textwidth}
\includegraphics*[width=0.9\textwidth,angle=0,clip]{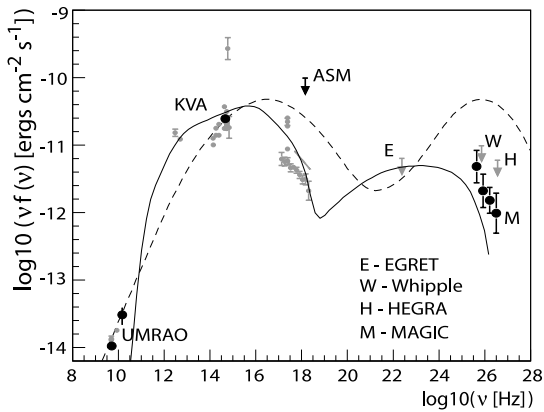}
\caption{\label{fig_SED} SED of   Mrk~180. Simultaneous data 
  (UMRAO, KVA, ASM, and MAGIC) are represented as black circles. 
  The gray circles represent historical data 
  \citep{giommi,perlman}.
  The arrows denote the upper limits from ASM, EGRET \citep{fichtel}, Whipple,
  and HEGRA.  The solid line is from \citet{costamante} and the dashed 
  line is from \citet{fossati} (see text).}
\end{minipage}
\end{center}
\end{figure}

Fig.~\ref{fig_SED} shows the SED of Mrk~180: the radio, optical and X-ray data
are from UMRAO, KVA, ASM, and the NED database, and the VHE data are from this analysis and
various upper limits.  Simultaneous data are noted in black, while historical
data are noted in gray.  The predicted flux by the phenomenological model of
\citet{fossati} ({\it dashed line}) is too high, as already emphasized by
the non detection measurements \citep{54AGN,horan}.  A more detailed SSC model
from \citet{costamante} ({\it solid line}) seems to describe the data more
closely.

The rather steep slope of the VHE spectrum suggests an IC peak position well
below 200~GeV, while the non detection by EGRET gives a lower limit of
$\sim1$~GeV for the peak position. This is in agreement with the SSC model of
\citet{costamante} suggesting that the IC peak position is at around 10~GeV.  The
overall IC luminosity, however, is underestimated in this model: the observed
integral flux above 300~GeV is a factor 30 larger than predicted.  An
explanation for this discrepancy could be the model's underlying assumption of
a quiescent-state synchrotron spectrum to obtain the IC flux.  This could
indeed suggest that our measurement was made during a high state. 

The discovery of VHE emission from Mrk~180 during an optical outburst makes it
very tempting to speculate about the connection between optical activity and
increased VHE emission.  Since Mrk~180 has not been observed prior to the
outburst with MAGIC and since the upper limits from other experiments are above the
observed flux level, further observations are needed.

{\textit{Acknowledgements:}}
We would like to thank the IAC for the excellent working conditions at the ORM
in La Palma. The support of the German BMBF and MPG, the Italian INFN, the
Spanish CICYT, ETH research grant TH~34/04~3, and the Polish MNiI grant
1P03D01028 is gratefully acknowledged.  We also thank H. D. and M. F. Aller,
at the University of Michigan Radio Observatory, for providing us with the UMRAO data.

\end{document}